\documentclass[12pt]{article}
\usepackage{graphicx}

\def\Re{{\cal R \mskip-4mu \lower.1ex \hbox{\it e}\,}}
\def\Im{{\cal I \mskip-5mu \lower.1ex \hbox{\it m}\,}}
\def\ie{{\it i.e.}}
\def\eg{{\it e.g.}}
\def\etc{{\it etc}}
\def\etal{{\it et al.}}

\def\sub#1{_{\lower.25ex\hbox{$\scriptstyle#1$}}}
\def\tev{\,{\ifmmode\mathrm {TeV}\else TeV\fi}}
\def\gev{\,{\ifmmode\mathrm {GeV}\else GeV\fi}}
\def\mev{\,{\ifmmode\mathrm {MeV}\else MeV\fi}}
\def\mpl{\ifmmode \overline M_{Pl}\else $\overline M_{Pl}$\fi}
\def\cc{\ifmmode k/\overline M_{Pl}\else $k/\overline M_{Pl}$\fi}
\def\lpi{\ifmmode \Lambda_\pi\else $\Lambda_\pi$\fi}
\def\to{\rightarrow}

\def\subw{_{\rm w}}
\def\mh{\ifmmode m\sbl H \else $m\sbl H$\fi}
\def\mch{\ifmmode m_{H^\pm} \else $m_{H^\pm}$\fi}
\def\mt{\ifmmode m_t\else $m_t$\fi}
\def\mc{\ifmmode m_c\else $m_c$\fi}
\def\mz{\ifmmode M_Z\else $M_Z$\fi}
\def\mw{\ifmmode M_W\else $M_W$\fi}
\def\mws{\ifmmode M_W^2 \else $M_W^2$\fi}
\def\mhs{\ifmmode m_H^2 \else $m_H^2$\fi}
\def\mzs{\ifmmode M_Z^2 \else $M_Z^2$\fi}
\def\mts{\ifmmode m_t^2 \else $m_t^2$\fi}
\def\mcs{\ifmmode m_c^2 \else $m_c^2$\fi}
\def\mchs{\ifmmode m_{H^\pm}^2 \else $m_{H^\pm}^2$\fi}
\def\ztwo{\ifmmode Z_2\else $Z_2$\fi}
\def\zone{\ifmmode Z_1\else $Z_1$\fi}
\def\mtwo{\ifmmode M_2\else $M_2$\fi}
\def\mone{\ifmmode M_1\else $M_1$\fi}
\def\tb{\ifmmode \tan\beta \else $\tan\beta$\fi}
\def\xw{\ifmmode x\subw\else $x\subw$\fi}
\def\ch{\ifmmode H^\pm \else $H^\pm$\fi}
\def\lum{\ifmmode {\cal L}\else ${\cal L}$\fi}
\def\inpb{\,{\ifmmode {\mathrm {pb}}^{-1}\else ${\mathrm 
{pb}}^{-1}$\fi}}
\def\infb{\,{\ifmmode {\mathrm {fb}}^{-1}\else ${\mathrm 
{fb}}^{-1}$\fi}}
\def\epem{\ifmmode e^+e^-\else $e^+e^-$\fi}
\def\ppb{\ifmmode \bar pp\else $\bar pp$\fi}
\def\bsg{\ifmmode B\to X_s\gamma\else $B\to X_s\gamma$\fi}
\def\bsll{\ifmmode B\to X_s\ell^+\ell^-\else $B\to X_s\ell^+\ell^-$\fi}
\def\bstt{\ifmmode B\to X_s\tau^+\tau^-\else $B\to X_s\tau^+\tau^-$\fi}
\def\lamt{\ifmmode \tilde\lambda\else $\tilde\lambda$\fi}
\def\shat{\ifmmode \hat s\else $\hat s$\fi}
\def\that{\ifmmode \hat t\else $\hat t$\fi}
\def\uhat{\ifmmode \hat u\else $\hat u$\fi}

\newskip\zatskip \zatskip=0pt plus0pt minus0pt
\def\matth{\mathsurround=0pt}
\def\lsim{\mathrel{\mathpalette\atversim<}}
\def\gsim{\mathrel{\mathpalette\atversim>}}
\def\atversim#1#2{\lower0.7ex\vbox{\baselineskip\zatskip\lineskip\zatskip
  \lineskiplimit 
0pt\ialign{$\matth#1\hfil##\hfil$\crcr#2\crcr\sim\crcr}}}

\def\grtsim{\,\,\rlap{\raise 3pt\hbox{$>$}}{\lower 
3pt\hbox{$\sim$}}\,\,}
\def\lsim{\,\,\rlap{\raise 3pt\hbox{$<$}}{\lower 3pt\hbox{$\sim$}}\,\,}

\def\grtsim{\,\,\rlap{\raise 3pt\hbox{$>$}}{\lower 3pt\hbox{$\sim$}}\,\,}
\def\lsim{\,\,\rlap{\raise 3pt\hbox{$<$}}{\lower 3pt\hbox{$\sim$}}\,\,}

\renewcommand{\thefootnote}{\fnsymbol{footnote}}

\begin{document} \begin{titlepage}
\rightline{\vbox{\halign{&#\hfil\cr
&SLAC-PUB-9731\cr
&May 2003\cr}}}
\begin{center}
\thispagestyle{empty}
\flushbottom

{\Large\bf Kaluza-Klein/$Z'$ Differentiation at the LHC and Linear Collider}
\footnote{Work supported by the Department of
Energy, Contract DE-AC03-76SF00515}
\medskip
\end{center}

\centerline{Thomas.G. Rizzo \footnote{e-mail:
rizzo@slac.stanford.edu}}
\vspace{8pt} \centerline{\it Stanford Linear
Accelerator Center, Stanford, CA, 94309}

\vspace*{0.7cm}

\begin{abstract}
We explore the capabilities of the LHC and the Linear Collider(LC) to 
distinguish the production of Kaluza-Klein(KK) excitations from an ordinary 
$Z'$ within the context of theories with TeV scale extra dimensions.  At 
the LHC, these states are directly produced in the Drell-Yan 
channel while at the LC the effects of their exchanges are indirectly felt as 
new contact interactions in processes such as $e^+e^-\to f\bar f$. While 
we demonstrate that the LC is somewhat more capable at KK/$Z'$ differentiation 
than is the LHC, the simplest LC analysis relies upon the LHC data for the 
resonance mass as an important necessary input. 
\end{abstract}

\renewcommand{\thefootnote}{\arabic{footnote}} \end{titlepage}

\section{Introduction and An Outline of the Problem}

The possibility of KK excitations of the Standard Model(SM) gauge bosons 
within the framework of theories with TeV-scale extra dimensions has been 
popular for some time{\cite {anton}}. The proven variety of such models 
is very large and continues to grow. For example, given 
the possibility of warped or flat extra dimensions one can  
construct a large number of interesting yet distinct models whose detailed 
structure depends upon a number of choices, \eg, whether all the gauge 
fields experience the same number of dimensions, whether the fermions and/or 
Higgs bosons are also in the bulk, whether brane kinetic terms are 
important{\cite{bkt}} in the determination of the KK spectrum and couplings 
and whether there exists a conservation law of KK number or KK parity, as in 
the case of the Universal Extra Dimensions(UED){\cite {ued}} scenario. 
If such KK gauge excitations do 
exist how will they be observed at colliders and how will we know that we 
have observed signals for extra dimensions and not some other new physics 
signature? For example, it is well known that UED might be mimicked 
by supersymmetry with somewhat degenerate superpartners 
at the LHC{\cite {ued}} unless the spins of the new KK states can be 
measured as can be done at the LC or the higher KK modes observed. 

In the analysis below we will be interested in the question of distinguishing 
the lightest KK excitations of the SM electroweak gauge bosons from a more 
conventional $Z'$ 
at both the LHC and LC. At the LHC, single KK/$Z'$ production is most easily 
observed via the Drell-Yan mechanism whereas, at the LC, the exchange of 
either set of states leads to contact interaction-like modifications to 
processes such as $e^+e^- \to f\bar f$. Here we will assume that the LHC 
discovers a single, rather heavy KK state whose mass is beyond the direct 
reach of the 
LC, a possibility consistent with the simplest TeV$^{-1}$ extra dimensions 
scenario. The state is assumed to be 
sufficiently massive, as indicated by current constraints 
from precision measurements,  so that higher KK states cannot be 
produced thus eliminating the most obvious signature for KK production  
Though other scenarios are of course possible, the one considered here is 
the simplest case to analyse; more complex possibilities will be studied 
elsewhere. 

The nature of our question and the above assumptions already limit our 
focus to a rather specific class of extradimensional 
theories and excludes many others. For example, at the tree level in UED, 
a conserved KK-parity exists which forbids the single production or exchange 
of KK 
states by zero modes and thus this class of theories is 
clearly excluded from our considerations. (We note, however, that at one loop 
the production of the {\it even} members of 
the KK tower are allowed by KK-parity 
conservation. The UED case and the standard $Z'$ scenario can then be 
most easily distinguished 
by the existence of the set of first KK excitations with masses essentially 
half that of the KK state observed in Drell-Yan.)  
In addition we can exclude 
models whose couplings and spectra are such that multiple KK resonances  
will be directly observable at the LHC. In this case there can be 
no issue of confusion as to whether or 
not extradimensional signatures are being produced (unless one is willing 
to postulate the existence of a conspiratorial multi-$Z'$ model).  
We also can exclude from consideration 
the set of models wherein the KK excitations 
of only the $SU(2)_L$ {\it or} 
$U(1)_Y$ gauge bosons can be produced. If either of 
these possibilities were realized {\it and} the spectrum of the KK fields was 
such that second and higher resonances were beyond the reach of the LHC, one 
can easily convince oneself that the KK and $Z'$ interpretations cannot be 
distinguished. (Of course the LC would tell us the couplings of these 
states and identify them as `copies' of those in the SM.) 
A similar situation holds for $W^\pm$ KK excitations    
even when the entire $SU(2)_L\times U(1)_Y$ gauge structure is in the bulk 
since there are only one set of KK excitations in this channel.   
Of course after making these few cuts in model space many theories remain 
to be examined and a full analysis of all the possibilities is beyond the 
scope of this paper. 

The simplest model of the class we will consider is the case of only one 
flat extra dimension; a generalization of our analysis to some of the 
more complex scenarios will be considered elsewhere.   
In this basic scheme all the fermions are constrained to  
lie at one of the two orbifold fixed points, $y=0,\pi R$, associated with the 
compactification on the orbifold 
$S^1/Z_2${\cite {bunch}}, where $R$ is the radius of 
the compactified extra dimension. Under usual circumstances a 3-brane is 
located at each of the fixed points upon which ordinary 4-d fields will 
reside. In principle, a SM fermion can be localized on the brane 
at either fixed point consistent with the constraints of 
gauge invariance. In our discussions below we will consider 
two specific cases: either all of the fermions are placed at $y=0$($D=0$), 
the standard situation, or 
the quarks and leptons are localized at opposite fixed points($D=\pi R$). 
Here $D$ is the distance between the quarks and leptons in the single 
extra dimension. 
The latter model, with oppositely localized quarks and leptons  
may be of interest in the suppression of proton decay in 
certain schemes. (Certainly more complicated scenarios are possible even if we 
assume generation independence and natural flavor conservation.) 
In such schemes the fermionic couplings of 
the KK excitations of a given gauge field are identical to those of the SM,  
apart from a possible sign if the relevant fermion under consideration 
lives at the $y=\pi R$ fixed point, and an overall factor of 
$\sqrt 2$. The gauge boson KK excitation masses are given, to lowest order in 
$(M_0/M_c)^2$, by the relationship $M_n^2=(nM_c)^2+M_0^2$, where $n$ labels 
the KK level, $M_c =1/R\sim 1$ TeV is the compactification 
scale and $M_0$ is the 
zero-mode mass obtained via spontaneous symmetry breaking for the cases of the 
$W$ and $Z$. Here we have assumed that any brane localized kinetic terms 
which may be present{\cite {bkt}} do not significantly alter these naive 
results.  Note that the first KK excitations of the photon and $Z$ 
will be highly 
degenerate in mass, becoming more so as $M_c$ increases. For example, if 
$M_c=4$ TeV the splitting between the first $Z$ and $\gamma$ KK states 
is less than 
$\sim$ 1 GeV, too small to be observed at the LHC. 

\begin{figure}[htbp]
\centerline{
\includegraphics[width=9cm,angle=90]{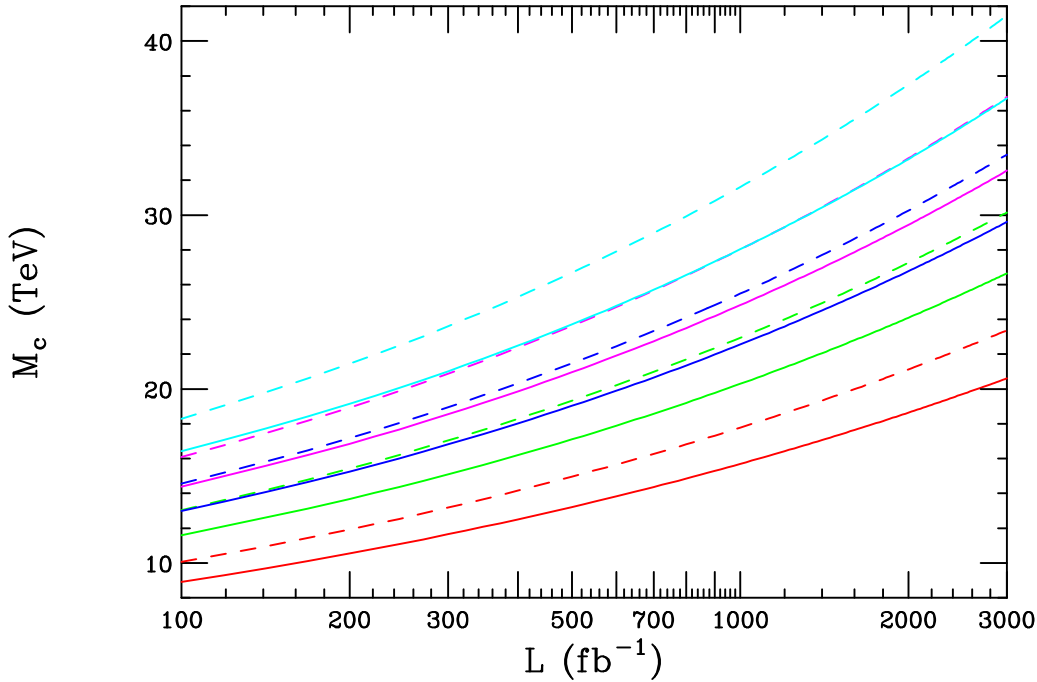}}
\vspace*{0.1cm}
\caption{$95\%$ CL bound on the scale $M_c$ as a function of the LC 
integrated luminosity from the reaction $e^+e^-\to f\bar f$, where 
$f=\mu, \tau, c,b,t$ have been summed over. The 
solid(dashed) curves assume a positron polarization $P_+=0(0.6)$; an 
electron polarization of $80\%$ has been assumed in all cases. From 
bottom to top the center of mass energy of the LC is taken to be 
0.5, 0.8, 1, 1,2 and 1.5 TeV, respectively.}
\label{fig0}
\end{figure}

An updated analysis{\cite {bunch}}  
of precision electroweak data implies that $M_c \gsim 4-5$ TeV, independently 
of whether the Higgs field vev is mostly in the bulk or on the brane or upon  
which of the fixed points the various SM fermions are confined. This is   
a mass range directly accessible to the LHC for resonance production in 
the Drell-Yan channel.   
Interestingly, at a LC with a center of mass energy of  
$\sqrt s= 500-1000$ GeV the effects of KK exchanges with masses 
well in excess of the $M_c\sim 4-5$ TeV 
range are also easily observable as is shown in Fig.~\ref{fig0}. This 
implies that there will be sufficient `resolving power' at an LC 
to examine the influence of somewhat smaller values of $M_c$ in detail. 

\begin{figure}[htbp]
\centerline{
\includegraphics[width=9cm,angle=90]{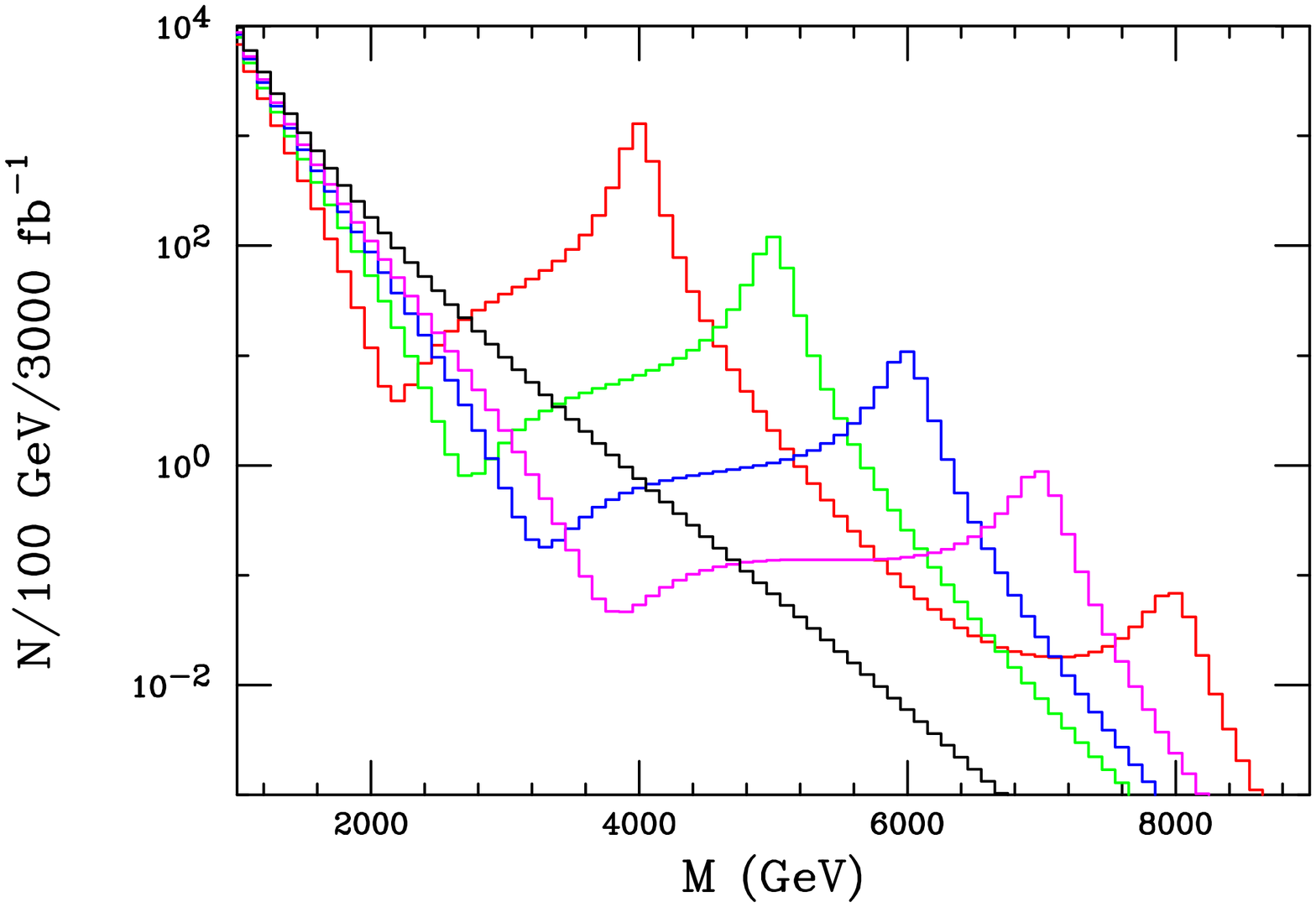}}
\vspace{0.4cm}
\centerline{
\includegraphics[width=9cm,angle=90]{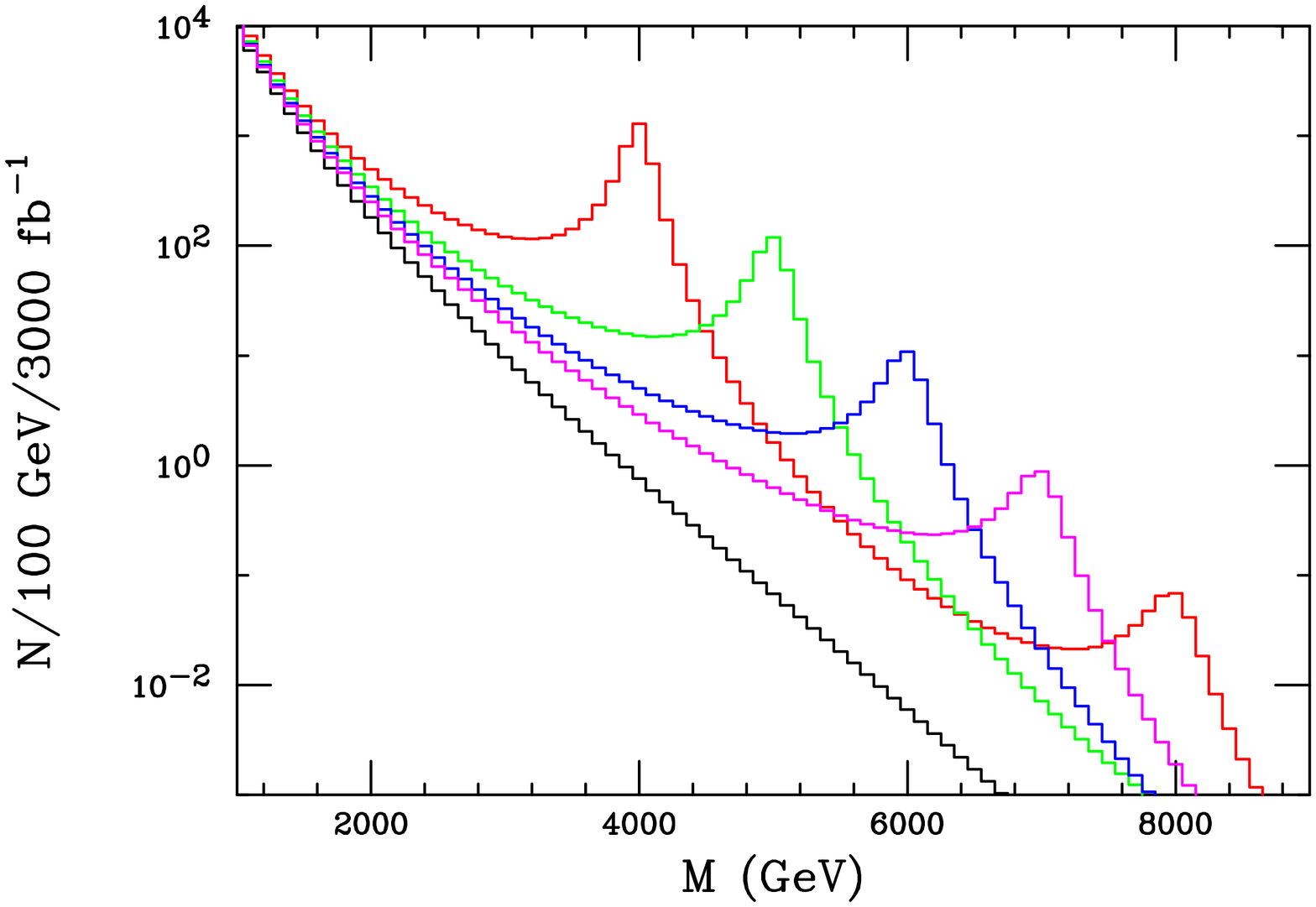}}
\vspace*{0.1cm}
\caption{Production rate, in the Drell-Yan channel $pp\to e^+e^- X$, 
for $\gamma/Z$ KK 
resonances as a function of dilepton invariant mass assuming a very high 
luminosity LHC. A rapidity cut $|\eta_l|\leq 2.5$ has been applied to the 
final state leptons. The red(green, blue, magenta) 
histogram corresponds $M_c$=4(5, 6, 7) TeV, respectively. The black histogram 
is the SM background. In the top panel all fermions 
are assumed to lie at the $y=0$ fixed point, $D=0$, while the quarks and 
leptons are split, $D=\pi R$,  in the lower panel.}
\label{figm1}
\end{figure}

Of course this large value of $M_c$ implies that 
the LHC experiments will at best observe only a {\it single} bump in the 
$\ell^+\ell^-$ channel as the next set of KK states, which are essentially 
twice as heavy, $\gsim 8-10$ TeV,  
are too massive to be seen even with an integrated luminosity of order 
$1-3$ $ab^{-1}${\cite {tgr}}. (Such high luminosities may be approachable 
at an upgrade of the LHC{\cite{upgrade}}; we will keep this possibility 
in mind in our analysis below.)    
This can be seen from Fig.~\ref{figm1} which follows from assumptions  
that all fermions lie at either the $y=0$ or $y=\pi R$ fixed points, \ie, 
$D=0$ or $\pi R$. These 
apparently isolated single resonance structures are, of course, 
superpositions  
of the individual excitations of both the SM $\gamma$ and $Z$ which are 
highly degenerate as we noted above. 
It is this dual excitation plus the existence of 
additional tower states that lead to the very unique resonance shapes that 
we see in 
either case.  Note that above the first KK resonance the excitation curves 
for the $D=0$ and $\pi R$ cases are essentially identical.   
This figure shows that KK states up to masses somewhat in excess  
of $\simeq 7$ TeV or so should 
be directly observable at the LHC or the LHC with a luminosity upgrade in a 
single lepton pair channel. 

It is important to note that if brane kinetic terms, we have so far ignored, 
are important then the bounds on $M_c$ from precision measurements  
can be significantly weaker due to reduced fermion-KK gauge 
couplings thus allowing for a much lighter first KK state. It may 
then be possible to directly observe the higher excitations so that no 
confusion with $Z'$ production would occur. However, parameter space regions 
may exist where such a possibility would not occur though the first 
excitation remains light; such scenarios are 
beyond the scope of the present analysis and will be considered elsewhere.


If a gauge KK resonance structure is observed in Drell-Yan, 
how will this observation be interpreted? Here we imagine a time line 
where the LC turns on after several years of data taking by the LHC at  
roughly the time of an LHC luminosity upgrade.   
Through straightforward measurement 
of the lepton pair angular distribution it will be known immediately that the 
resonance is spin-1 and not, \eg, a spin-2 graviton resonance as in the 
Randall-Sundrum{\cite {rs}} model{\cite {dhr}}, provided sufficient 
luminosity is available. In addition, the existence 
of an essentially degenerate pair of resonances in both the charged and 
neutral Drell-Yan channels will forbid a possible graviton  interpretation. 
Perhaps the most straightforward possibility for interpretation would be that 
of an extended gauge model{\cite {snow}} which predicts the 
existence of a degenerate $W'$ and $Z'$; many such models already exist in the 
literature{\cite {models}}. Is it possible to distinguish this 
degenerate $Z'/W'$ model from KK excitations without seeing any of the rest of 
the tower states? Clearly, based on the discussion above, we must 
focus on differentiating the $Z'$ from the first (and only observable) 
KK excitation spectra below and around the peak. 
At least temporarily, only LHC data will be available for this discrimination 
until the LC subsequently turns on.

\section{What Can the LHC Tell Us?}

The issue of KK/$Z'$ differentiation at the LHC has been previously discussed 
to a limited extent by several authors in Ref.{\cite {before}}. The purpose 
of this section is to generalize those analyses as well as to extend them 
to the case of higher integrated luminosities. Also, we need to eventually 
make comparisons of the LHC results with those obtainable from the LC. 
Though hopefully more comprehensive than this 
earlier work, the present analysis will still leave much that 
remains to be studied even for the rather simple model we choose to 
examine here. Pictorially we will consider the 
case $M_c=4$ TeV but our analysis will be extended to significantly larger 
compactification mass values. 

Fig.~\ref{fig2} shows a closeup of 
the excitation spectra and forward-backward asymmetries, $A_{FB}$,  
for KK production near the first resonance region 
assuming $M_c=4$ TeV and with $D=0,\pi R$. There are 
several comments to be made at this point before we begin our analysis. 
First, for $pp$ colliders, note that the 
forward-backward asymmetry is defined via the 
angle made by the direction of the negatively charged lepton and the 
direction of motion of the center of mass in the laboratory frame. 
This direction is assumed to be the same as that of the initial state quark, 
which is reasonable given the harder valence parton distribution. 
Second, we observe the by now familiar strong 
destructive interference minimum{\cite {anton} in the cross section for the 
$D=0$ case near $M\simeq 0.55M_c$ which is also reflected in the 
corresponding narrow 
dip in the asymmetry. This dip structure is a common feature that will 
persist even in higher dimensional models or in models with warped extra 
dimensions. The precise location of the dip is sensitive to model details, 
however. 
Third, we notice that the overall behaviour of the $D=0$ and $D=\pi R$ 
cases is completely different below the peak while almost identical 
above it. In fact, if anything, the $D=\pi R$ case 
displays a strong {\it constructive} interference in the region below the 
KK peak. This difference in the two excitation curves is due solely to the 
additional factor of $(-1)^n$ appearing in the KK sum arising from 
the placement of  
the quarks and leptons at opposite fixed points. (Here, $n$ labels the KK 
number of the state.) Lastly, we note that the 
peak cross section and peak $A_{FB}$ values, respectively, are nearly 
identical in the two cases. 
In the narrow width approximation we find that 
the two sets of values are identical 
since the additional sign factors cancel. 

\begin{figure}[htbp]
\centerline{
\includegraphics[width=9cm,angle=90]{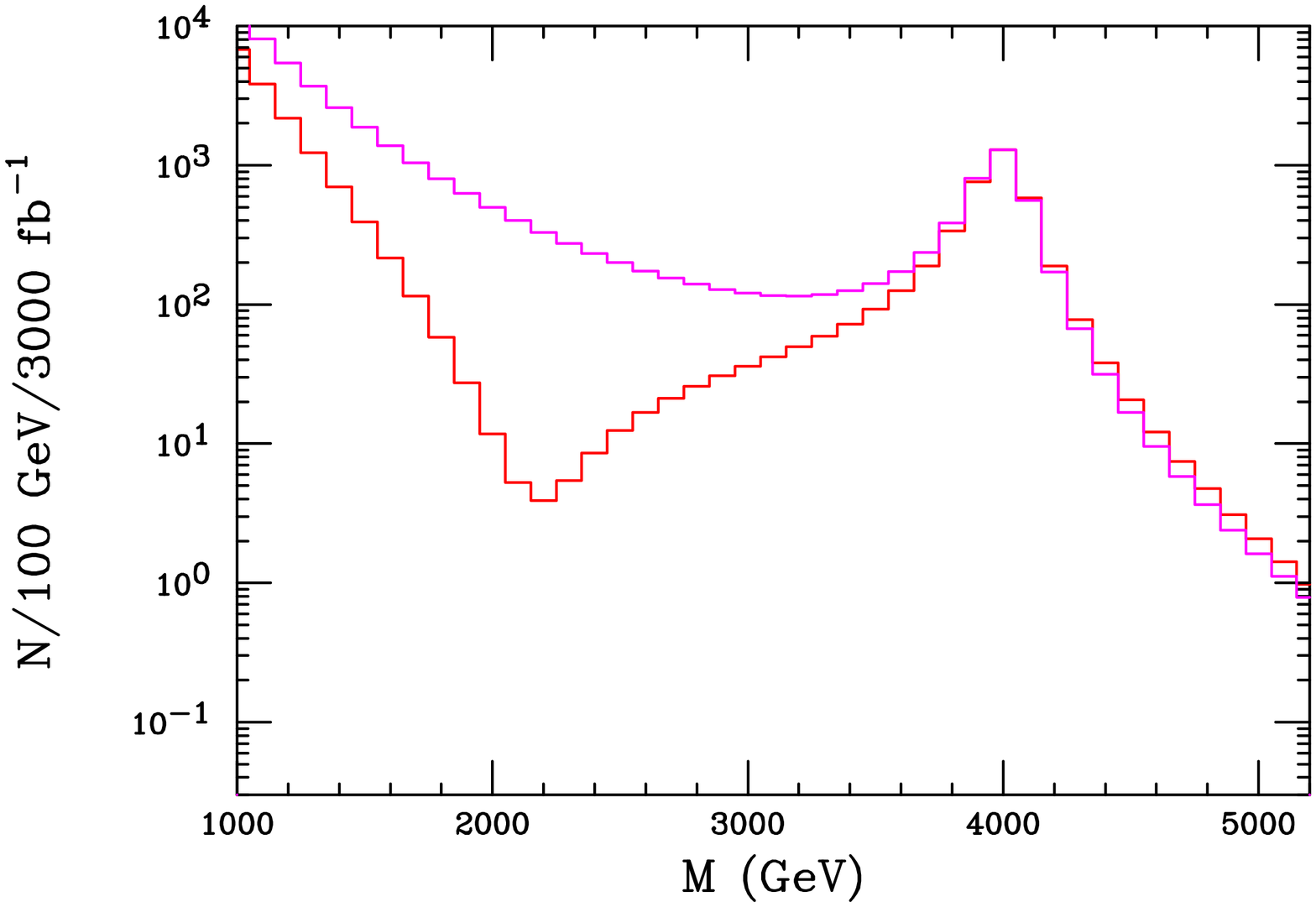}}
\vspace*{0.4cm}
\centerline{
\includegraphics[width=9cm,angle=90]{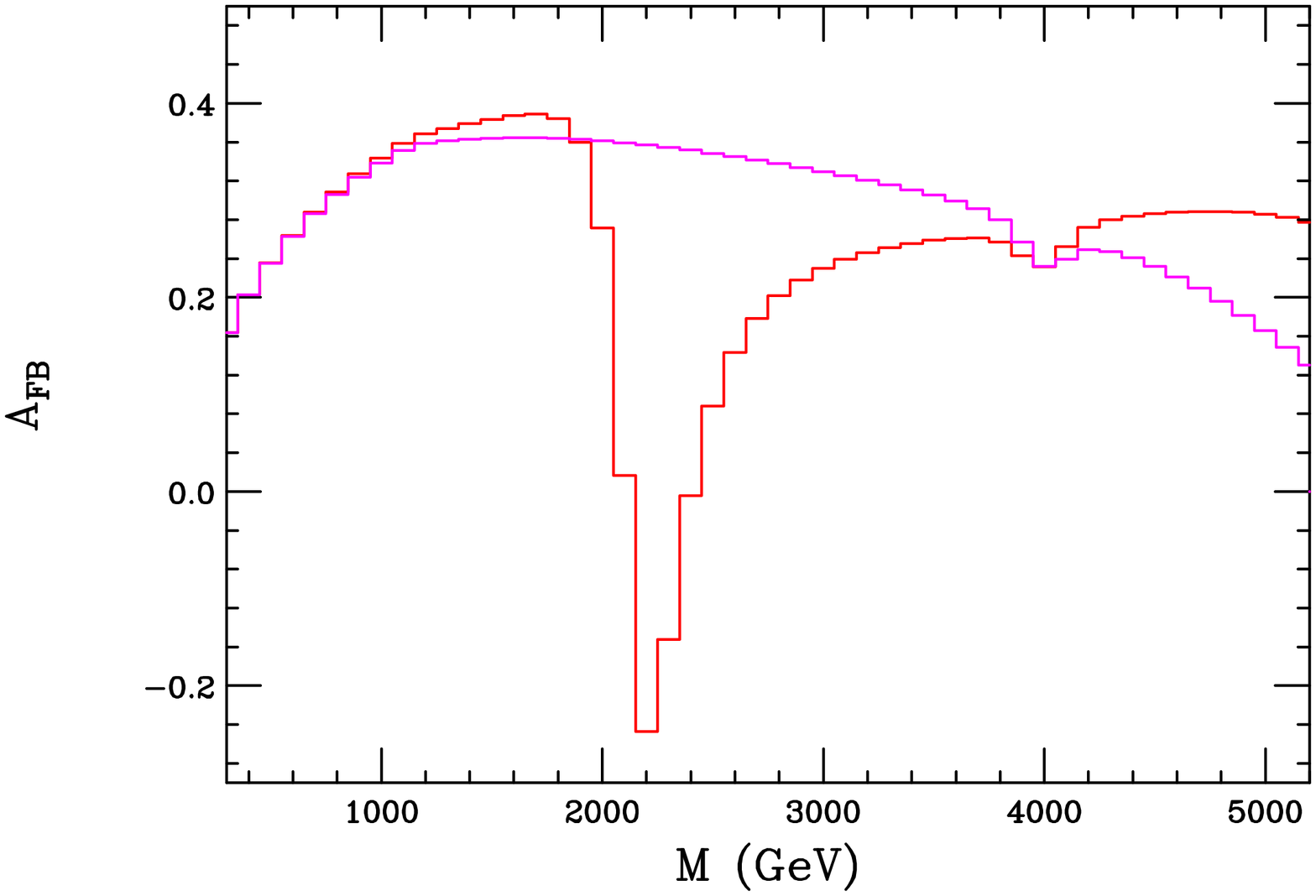}}
\vspace*{0.1cm}
\caption{A comparison of the lepton pair invariant mass spectrum and 
forward-backward lepton asymmetry for the production of a 4 TeV KK 
resonance for the two choices of $D$. The red(magenta) histograms are for 
the case $D=0(\pi R)$.}
\label{fig2}
\end{figure}

For either $D$ choice the excitation curve and $A_{FB}$ appear to be 
qualitatively  different than that which one obtains for typical $Z'$ 
models{\cite{snow}} as is shown in Figs.~\ref{fig3} and \ref{fig32}. None 
of the dozen $Z'$ models produces resonance structures that appear anything 
like that seen for the KK case. The resonance structure 
for the KK case is significantly wider and has a larger peak cross section 
than does the typical $Z'$ model where the strong destructive interference 
below the resonance is absent. (We remember however that the height and 
width of the $Z'$ or KK resonance also depends on the set of allowed decay 
modes.)  
In addition, the dip in the value of $A_{FB}$ occurs much closer to the 
resonance region for the typical $Z'$ model than it does in the KK case. 
Clearly, while the KK resonance does not look like one of the usual $Z'$'s,  
we certainly could not claim, based on these figures, that some 
$Z'$ model with which we are not familiar 
cannot mimic either KK case. In fact, from the figures, one can more easily 
imagine a $Z'$ with stronger than typical couplings leading to an excitation 
structure similar to the $D=\pi R$ KK case, \ie, it seems more likely that the 
case of $D=\pi R$ can be mimicked by a (strongly coupled) $Z'$ than 
does the $D=0$ case.

\begin{figure}[htbp]
\centerline{
\includegraphics[width=9cm,angle=90]{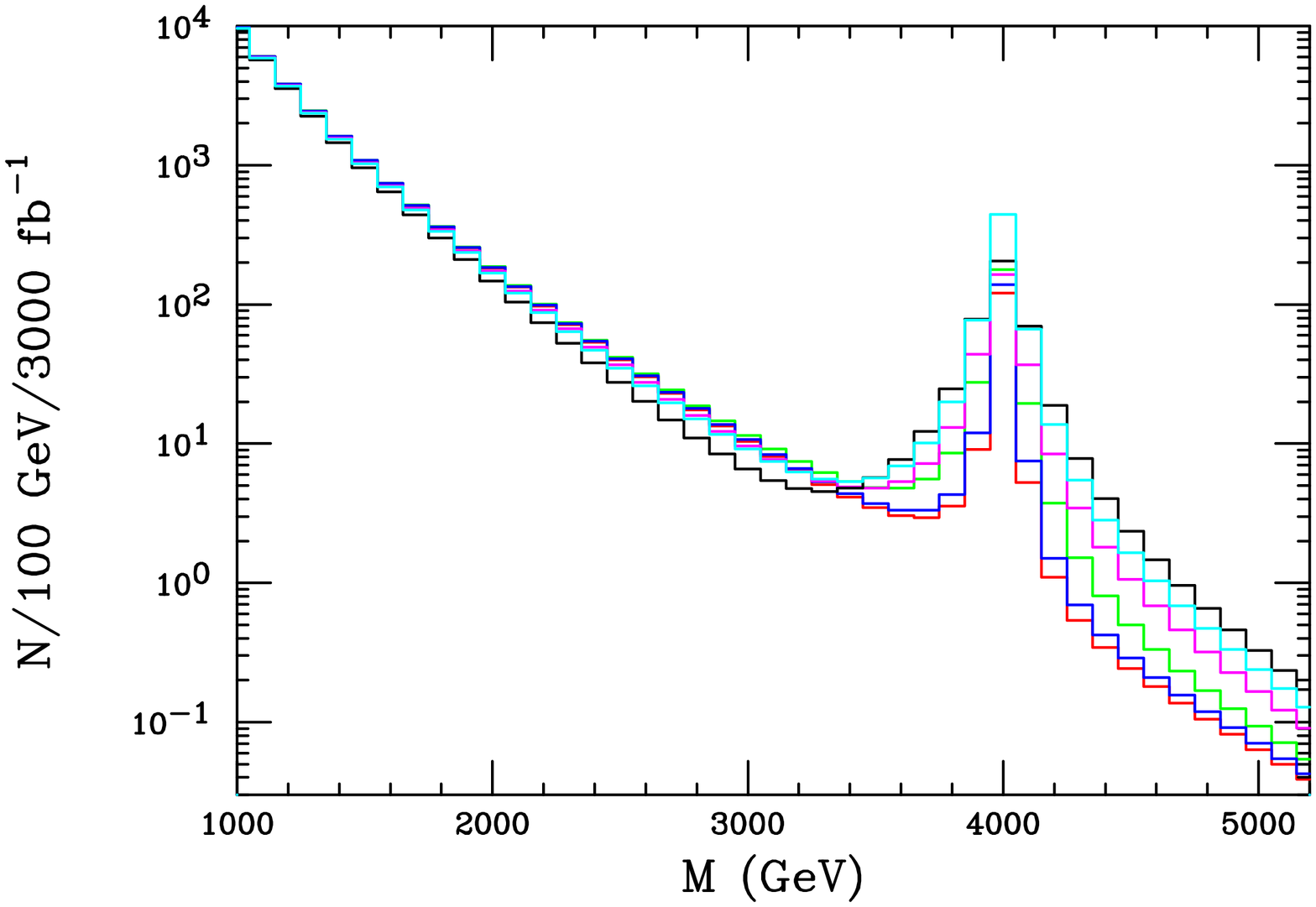}}
\vspace*{0.4cm}
\centerline{
\includegraphics[width=9cm,angle=90]{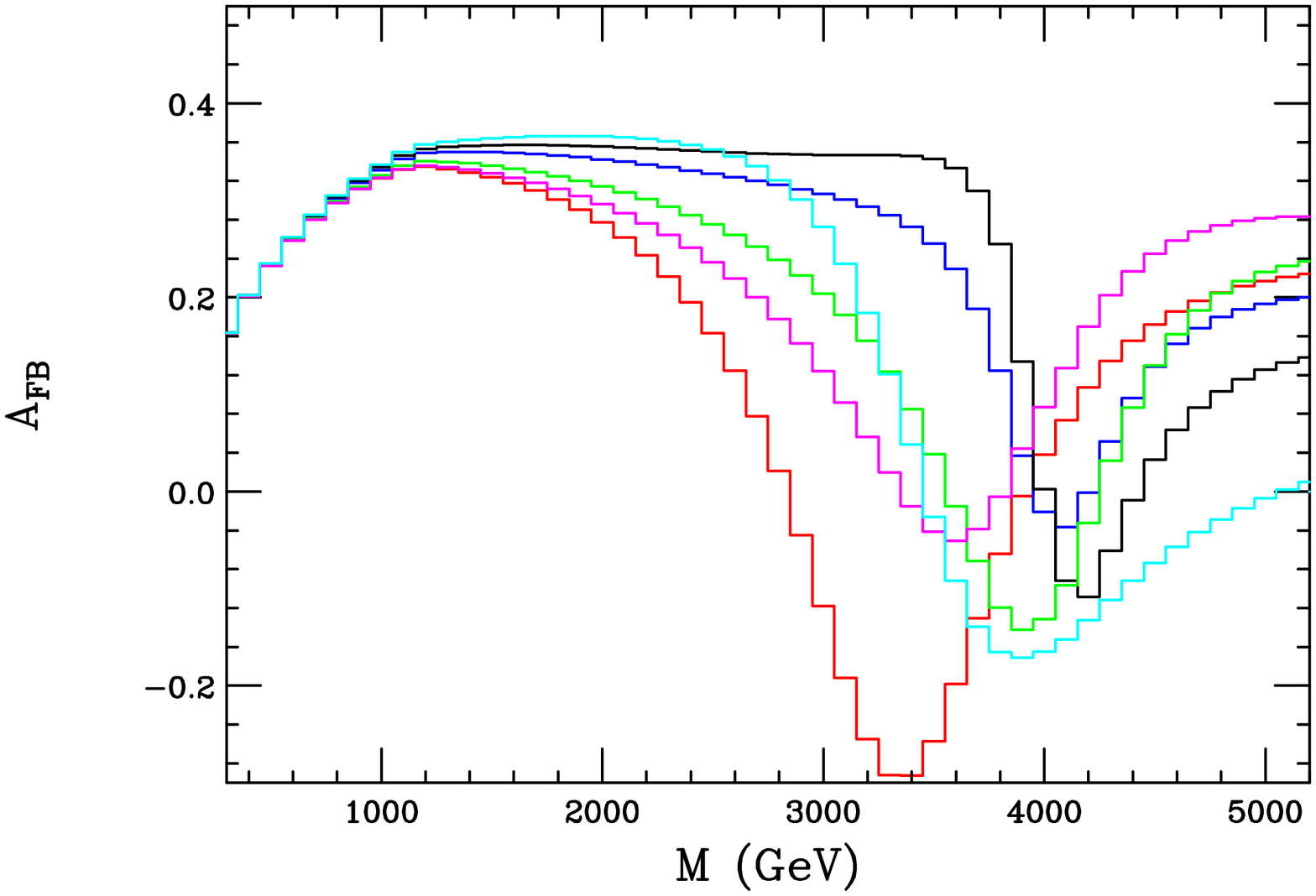}}
\vspace*{0.1cm}
\caption{Same as in the previous pair of figures but now for a number of 
$Z'$ models. The red(green, blue, magenta, cyan, black) 
histograms correspond to $E_6$ model $\psi(\chi, ~\eta)$, the Left Right 
Symmetric Model with $\kappa=g_R/g_L=1$, the Alternative Left Right Model and 
the Sequential Standard Model, respectively. For descriptions of these 
models and original references see Ref.{\cite {snow}}.}
\label{fig3}
\end{figure}
\begin{figure}[htbp]
\centerline{
\includegraphics[width=9cm,angle=90]{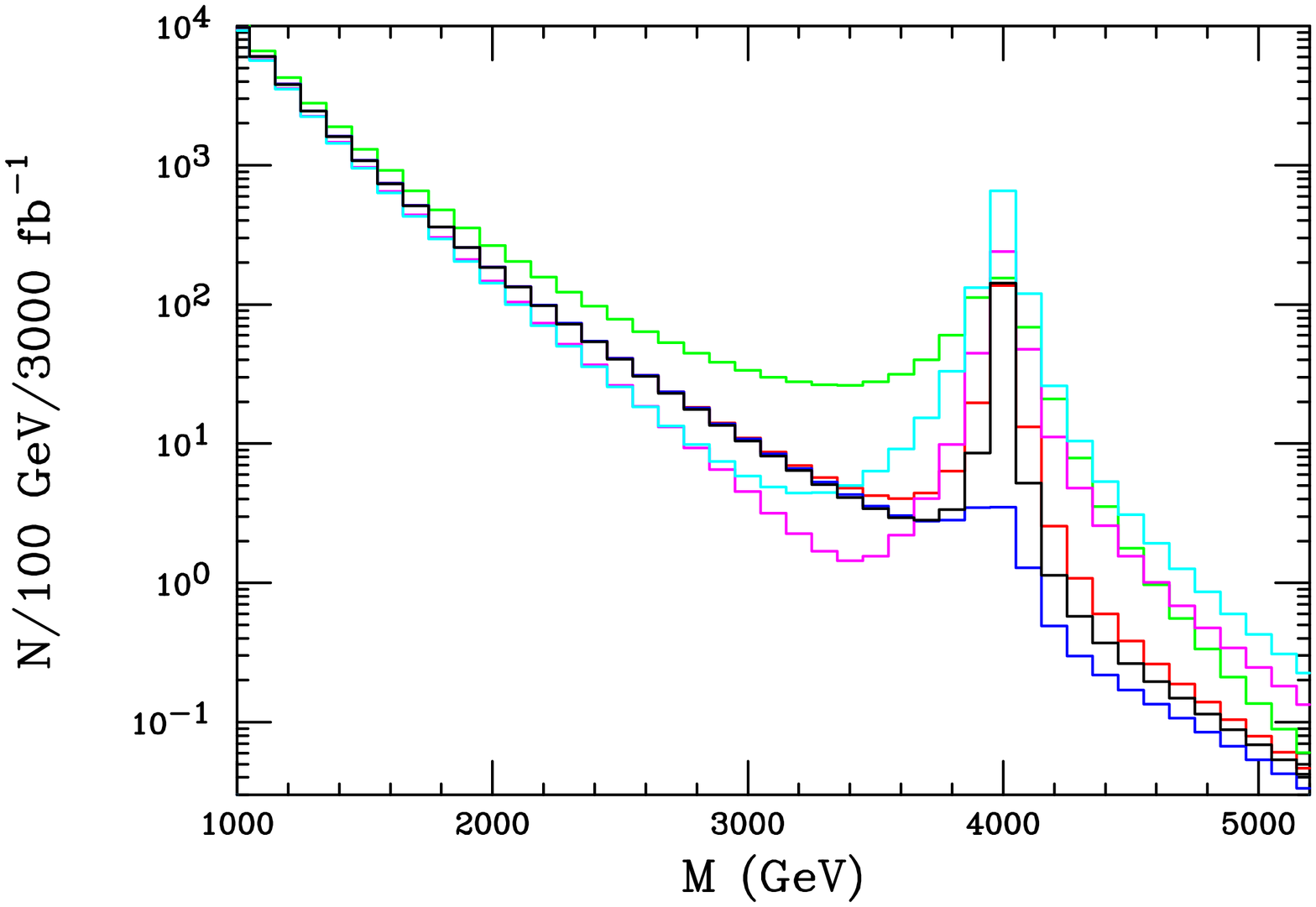}}
\vspace*{0.4cm}
\centerline{
\includegraphics[width=9cm,angle=90]{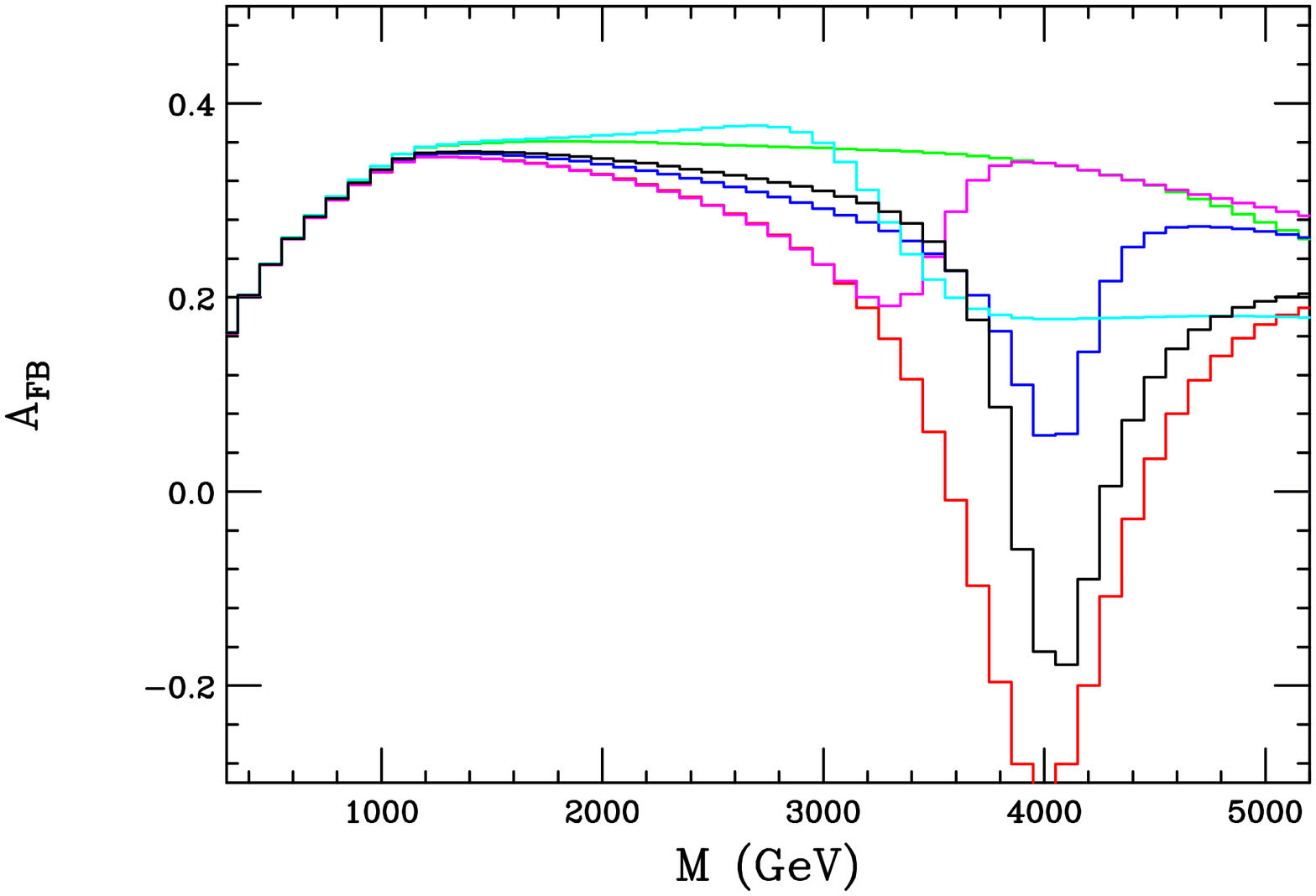}}
\vspace*{0.1cm}
\caption{Same as in the previous pair of figures but now for a further set 
of $Z'$ models. The red(green, blue, magenta, cyan, black) 
histograms correspond to $E_6$ model $I$, the Un-unified Model with 
$s_\phi=0.6$, the Foot-Hernandez model, the model of Kuo \etal, a $Z'$ 
coupling proportional to hypercharge and $E_6$ model $\eta$ with gauge 
kinetic mixing parameter $\delta=0.25$, respectively. For descriptions 
of these models and original references see Ref.{\cite {snow}}.}
\label{fig32}
\end{figure}

In order to quantify the differences between the KK and $Z'$ 
scenarios we must choose observables that have reasonable statistical 
power associated with them and do 
not explicitly depend on any assumptions about how the KK or $Z'$ may 
decay, \eg, if the resonance has non-SM decays such as into supersymmetric 
final states. 
Consider the $D=0$ case; given the $100-3000$ $fb^{-1}$ luminosity of 
the LHC and its upgrade, the 
invariant mass distribution will only be useful as an observable for lepton 
pair masses above the $Z$ pole and below $\sim 2.5$ TeV and as well as 
near the KK/$Z'$ resonance. Outside these regions there is either no 
sensitivity to new physics or the event rate is just too low to make 
decent measurements. As we noted above the resonance region 
we cannot use fairly if we assume that all of the decays of the resonance are 
a priori unknown. Once we are safely  
beyond the peak region the cross section 
is quite small yielding too poor a set of statistics to be valuable.
 
Since the statistics required to determine $A_{FB}$ is 
significantly higher for a fixed value of the invariant mass than it is 
for the mass distribution itself  
(because angular distributions now need to be measured as 
well), its range of usefulness as a differentiating observable is 
more restricted. Perhaps below $\sim 1.5$ TeV in dilepton invariant mass 
sufficient statistics will be available to allow $A_{FB}$ to be useful. 
However, one might imagine that $A_{FB}$ may also be helpful very near 
the peak region since we know that for the case of a $Z'$, $A_{FB}$  
is approximately independent of what modes the resonance may decay 
into, unlike the cross section. Naively, we would expect the inclusion 
of input from $A_{FB}$ data on the peak to improve the results 
obtained below.  However, it has been shown in our 
earlier work{\cite {before}} that even near the apparent KK 
pole, $A_{FB}$ depends on the relative total widths of the individual 
$\gamma$ and $Z$ KK excitations, which is model sensitive. 
In the analysis presented below we will ignore 
any additional guidance that may arise from considering the values of 
$A_{FB}$ at lepton pair masses below $\sim 1.5$ TeV 
and examine only the invariant mass distribution, \ie, the 
possible additional information 
obtainable from $A_{FB}$ will be ignored in the present analysis and will 
be left for later study.

Turning to our analysis, for $M_c=4(5,~6)$ TeV 
we begin by generating cross section `data' corresponding to dilepton masses 
in the range 
250-1850(2150, 2450) GeV in 100 GeV bins for both the $D=0$ and $\pi R$ cases. 
To go any lower in mass would not be very useful as we are then 
dominated by either 
the $Z$ peak or the photon pole. For larger masses the cross section is either 
too small in the $D=0$ case or is dominated by the heavy resonance as 
discussed above.  Next we try to fit these cross section 
distributions by making the assumption that the data 
is generated by a single $Z'$. For simplicity, we restrict our attention to 
the class of $Z'$ models with generation-independent couplings and where 
the $Z'$ has associated with it a new gauge group 
generator that commutes with weak isospin. 
These conditions are satisfied, \eg,  by GUT-inspired $Z'$ models 
as well as by many others in the literature{\cite {snow}}.
If these constraints hold then the $Z'$ couplings 
to all SM fermions can be described by only 5 independent parameters: the 
couplings to the left-handed quark and lepton doublets and the corresponding 
ones to the right-handed quarks and leptons. 
We then vary all of these couplings independently  
in order to obtain the best $\chi^2/df$ fit to the dilepton mass 
distribution and obtain the 
relevant probability/confidence level(CL) using statistical errors 
only. Systematic errors arising from, \eg, parton distribution function 
uncertainties will be ignored. In practice this is a  
fine-grained scan over a rather large volume of the 5-d parameter space  
examining more than $10^{10}$ coupling combinations for each of the cases we 
consider to obtain the best probability. 
In performing this fit it is assumed that the apparent $Z'$ mass is the 
same as that of the produced KK state which will be directly measured. 
In this approach, the overall 
normalization of the cross section is determined at the $Z$-pole which is 
outside of the fit region and is governed solely by SM physics.

\begin{figure}[htbp]
\centerline{
\includegraphics[width=9cm,angle=90]{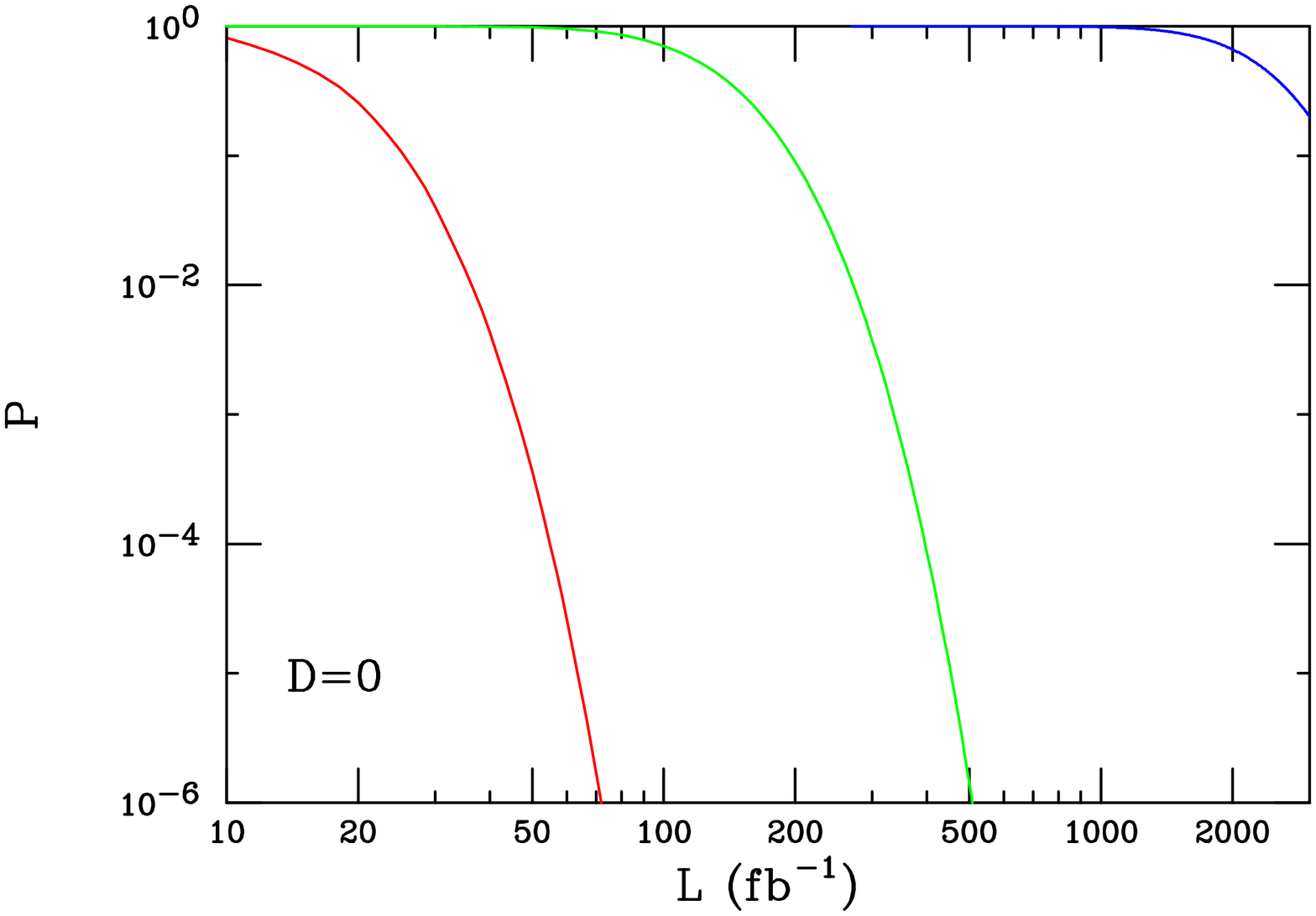}}
\vspace*{0.4cm}
\centerline{
\includegraphics[width=9cm,angle=90]{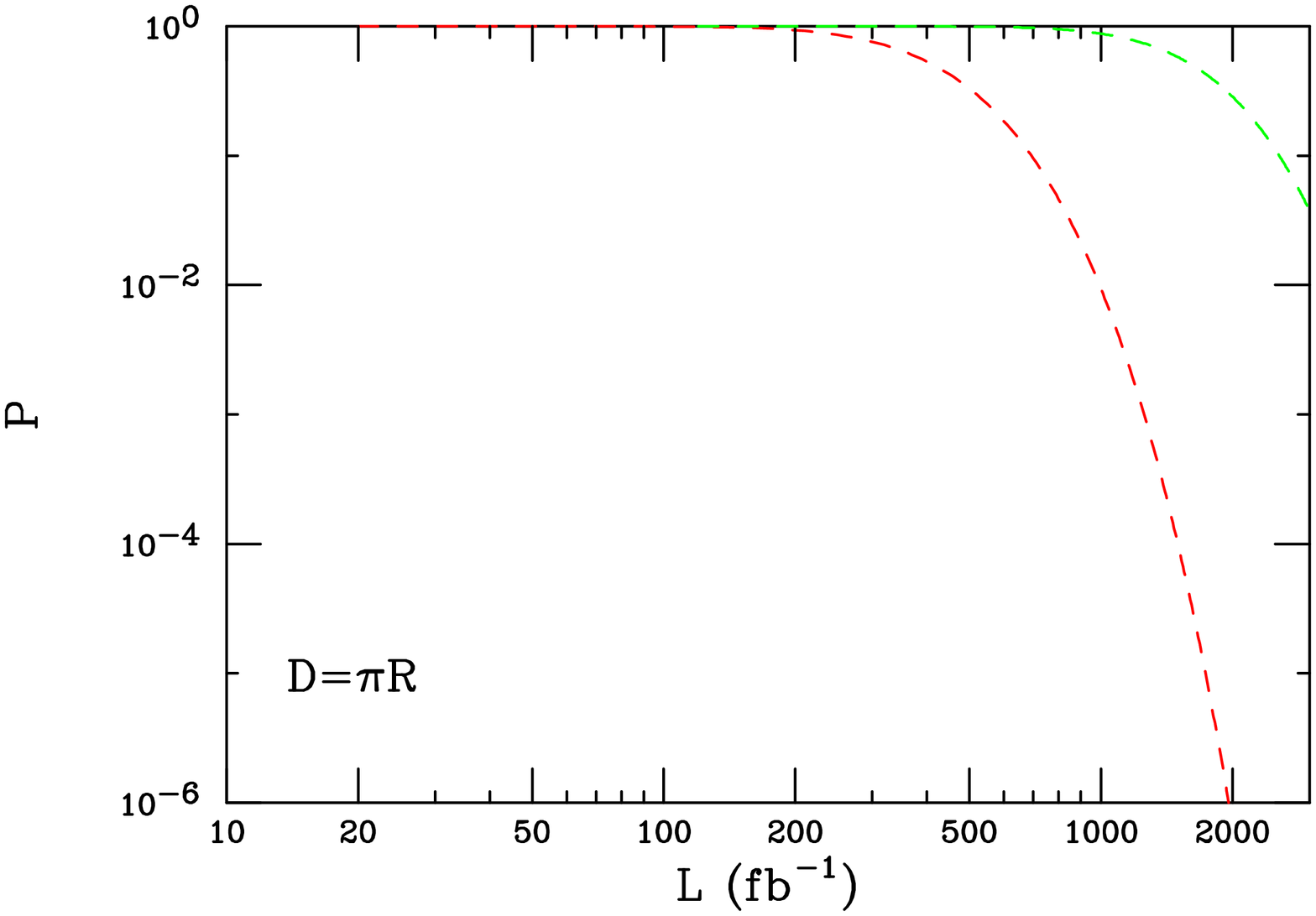}}
\vspace*{0.1cm}
\caption{Probability associated with the best $Z'$ fit hypothesis as a 
function of the LHC integrated luminosity for the cases $D=0$ and $D=\pi R$. 
From left to right the curves correspond to the choice $M_c=4$, 5 and 6 
TeV, respectively.}
\label{fig4}
\end{figure}

The results of performing these fits for different values of $M_c$ and the 
two choices $D=0,\pi R$ are shown in Fig.~\ref{fig4}. Explicitly, these show 
the best fit probability for the $Z'$ hypothesis to the KK generated 
data. For example, taking the case $D=0$ with $M_c=4$ TeV we see that with an 
integrated luminosity of order 60 $fb^{-1}$ the best fit probability is 
near a few $\times 10^{-5}$. For such low probabilities we can certainly 
claim that the KK generated `data' is not well fit by the $Z'$ hypothesis. 
As the mass of the KK state increases the size of the shift in the production 
cross section from the SM expectation is reduced and greater statistics are 
needed to obtain the same probability level. For $M_c=5$ TeV we see that an 
integrated luminosity of order 400 $fb^{-1}$ is required to get to the same 
level of rejection of the $Z'$ hypothesis as above. Similarly, for $M_c=6$ TeV 
extremely high luminosities of order 7-8 $ab^{-1}$ would be required to get 
to this level of probability, which is most likely a factor of two or so 
beyond even that expected for the LHC upgrades unless combined data from 
both detectors was used.  

For the $D=\pi R$ case we see the situation is somewhat different in that 
the level of `confusion' between the KK and $Z'$ is potentially greater. 
This is what we might have expected based on our discussion above. Even for 
the case $M_c=4$ TeV we see that only at very high integrated luminosities 
of order $\sim 1.5~ab^{-1}$  can the KK and $Z'$ scenarios be distinguished 
at the level discussed above. 
With $M_c=5$ TeV, approximately 6 $ab^{-1}$ would be required to reach the  
same rejection level. For larger KK masses this separation becomes 
essentially impossible at the LHC.

\section{What Can the LC Tell Us?}

The analysis for the LC is somewhat different than at the LHC. No actual 
resonances are produced but deviations from SM cross sections and 
asymmetries are observed due the $s$-channel exchanges of the 
$Z'$ or KK gauge boson towers. Though subtle these two sets of 
deviations are not identical and our hope here 
is to use the precision measurement capability of a LC to distinguish them. 
We will assume that data is taken at a single value of $\sqrt s$ so that 
the mass of the KK or $Z'$ resonance obtained from the LHC 
must be used as an input to the analysis 
as presented here. Without such an input the analysis below can still be 
performed provided data from at least two distinct values of $\sqrt s$ are 
used as input{\cite {oldt}}. In that case $M_c$ becomes an additional fit 
parameter to be determined by the analysis from the $\sqrt s$ dependence 
of the deviations from the SM expectations. While this more general 
situation is certainly very 
interesting it is beyond the scope of the present analysis.

\begin{figure}[htbp]
\centerline{
\includegraphics[width=9cm,angle=90]{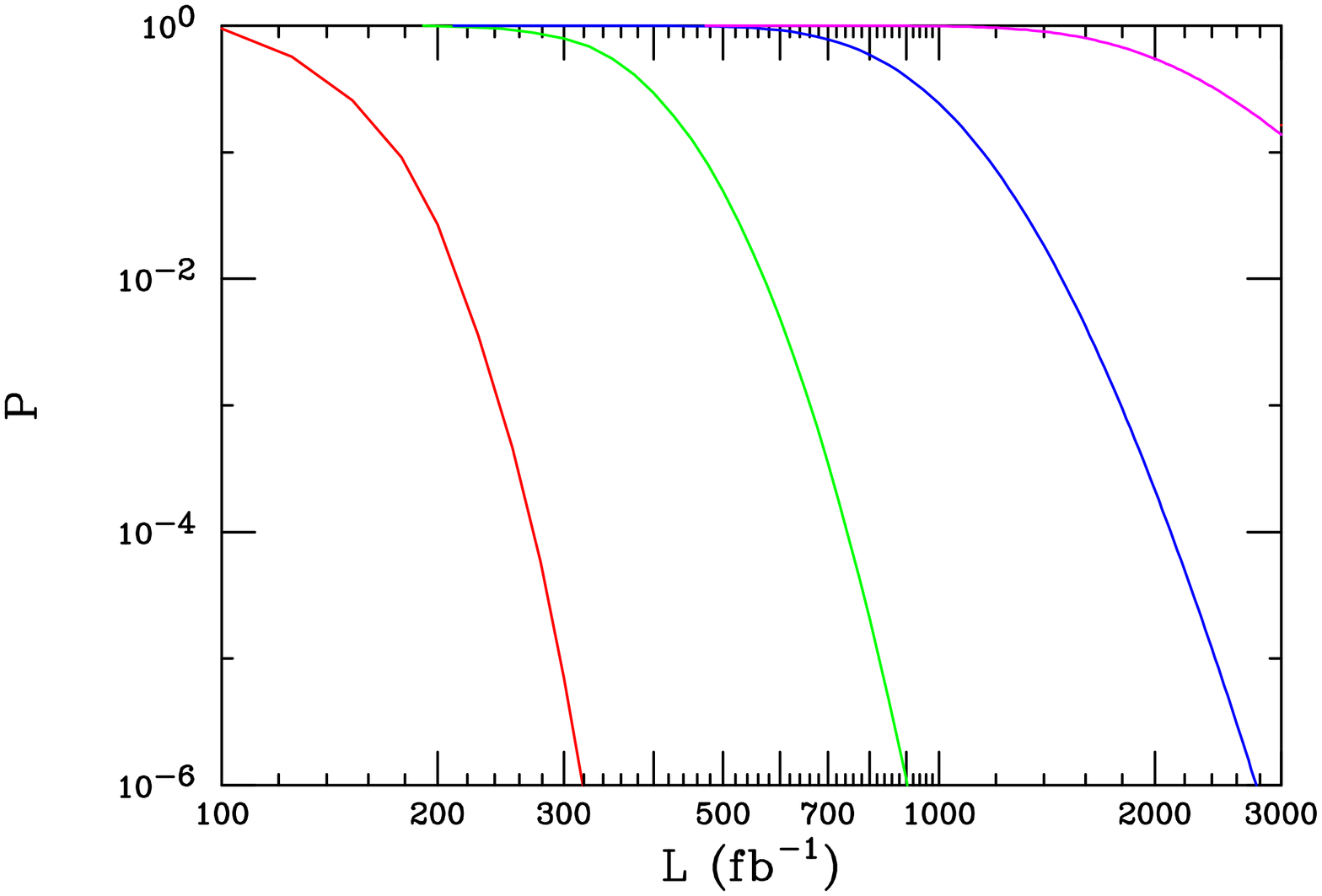}}
\vspace*{0.4cm}
\centerline{
\includegraphics[width=9cm,angle=90]{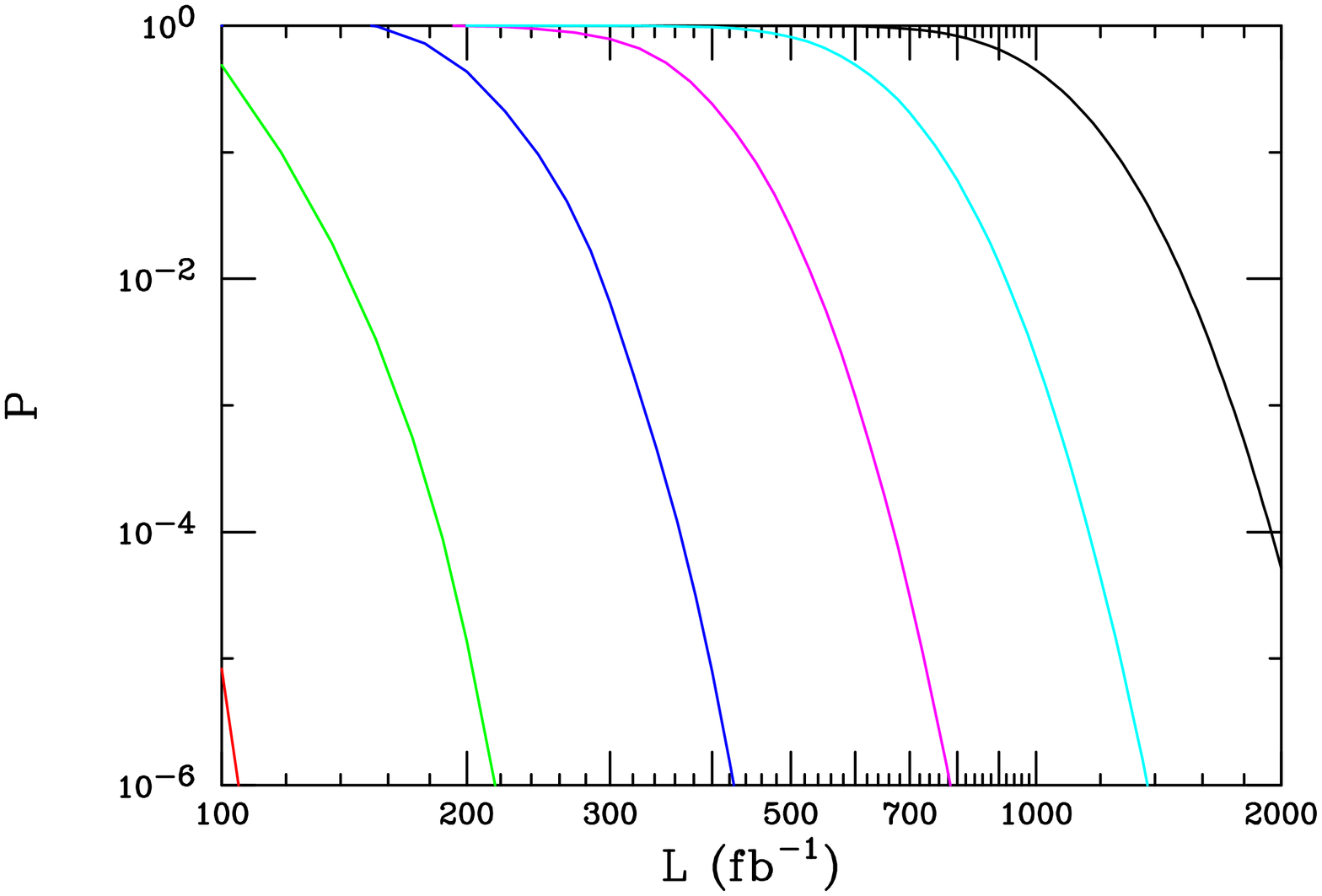}}
\vspace*{0.1cm}
\caption{Same as the previous figure but now for a 500(1000) GeV LC in the 
top(bottom) panel. The $D=0$ and $D=\pi R$ cases are identical here. From 
left to right the curves are for the cases $M_c=4,5,6,..$ TeV \etc. The 
value of $M_c$ is assumed to be determined at the LHC.}
\label{fig5}
\end{figure}

Consider the general process $e^+e^-\to f\bar f$; assuming KK states are 
actually present with a fixed $M_c$ as above, we generate `data' for both 
the differential cross section as well as the Left-Right polarization 
asymmetry, $A_{LR}$, including the effects of ISR, as functions of the 
scattering angle, \ie, $\cos \theta$, in 20 (essentially) equal sized bins. 
The electron beam is assumed to be $80\%$ polarized and angular acceptance 
cuts are applied.
Our other detailed assumptions in performing this analysis are the same 
as those employed in earlier studies and  
can be found in Ref.{\cite {oldtt}}. 
We then attempt to fit this `data' making the assumption 
that the deviations from the SM are due to a single $Z'$.  
For simplicity, here we will
concentrate on the processes $e^+e^- \to \mu^+\mu^-,\tau^+\tau^-$ as only 
the two leptonic couplings are involved in performing any fits. In this case 
the $D=0$ and $D=\pi R$ scenarios lead to {\it identical} results for the 
shifts in all observables at the LC, an advantage over the LHC case. 
Adding new  
final states, such as $b\bar b$ or $c\bar c$, may lead to potential 
improvements although additional fit parameters now must be introduced and 
the $D=0$ and $D=\pi R$ predictions would then again be distinct as at 
the LHC. To be specific, 
we consider two cases for the LC center of mass energy: $\sqrt s=0.5$ and 1 
TeV. We remind the reader that while these two leptonic couplings are 
adequate to describe the effects of $Z'$ exchange the description fails in 
the case of KK states since towers of both the $\gamma$ and $Z$ are being 
exchanged; this naturally requires three couplings in general since the 
photon tower has only vector couplings to SM fermions. 

As before in the LHC case we next vary the two assumed $Z'$ couplings to 
leptons to obtain the 
best $\chi^2/df$ for the fit which then leads to the probabilities shown in 
Fig.~\ref{fig5}. For the case of a $\sqrt s$=500 GeV LC, we see that an 
integrated luminosity of 300 $fb^{-1}$ is roughly equivalent to  
60 $fb^{-1}$ 
at the LHC for the case of $M_c=4$ TeV assuming $D=0$. 
For larger values of $M_c$ the 500 GeV LC does slightly 
better at KK/$Z'$ discrimination than the LHC: 
800(2200) $fb^{-1}$ at the LC equivalent is found to be roughly equivalent 
to 400(7500) $fb^{-1}$ at the LHC 
assuming $M_c=5(6)$ TeV. Since the $D=0$ and $\pi R$ cases are 
identical at the LC a further advantage is obtained there as noted earlier. 
Once the LC energy increases to 1 TeV the LC is seen to be superior in model 
separation but the analysis still relies upon the LHC to input the 
value of $M_c$ in 
the fits. The lower panel in Fig.~\ref{fig5} shows results for values 
of $M_c$ beyond the range of 7-8 TeV which is directly observable at the LHC. 
This would seem to imply that 
by extending the present analysis to include input from at least two values 
of $\sqrt s$ we may be able to extend the KK/$Z'$ separation out to very 
large masses at the LC.

As a final point one may wonder about the reverse problem, \ie, if the 
heavy resonance observed at the LHC {is} a $Z'$, how do we rule out the 
possibility of it being a KK state? From performing the analysis 
discussed above at both the LHC and LC we would know that the resonance 
couplings would be consistent with being a $Z'$ and not a KK state (at least 
not one of the type we have considered here). In particular, given the mass 
of the state from the LHC, the LC with its excellent $b$ and $c$ tagging 
capabilities would be able 
to provide a good fit to the various flavor couplings of the $Z'$ in addition
to the leptonic ones considered here. Thus it seems reasonably 
straightforward  that if a $Z'$ is 
discovered, given the present analysis 
and its extensions to other final states at the LC, it will be quite 
clear that it {\it is} indeed a $Z'$ and not a KK state.

\section{Summary and Conclusion}

New physics signatures arising from different sources may be confused when 
first observed at future colliders. Thus it is important to examine how 
various scenarios may be differentiated given the availability of only 
limited information. 
In this analysis we have performed a comparison of the capabilities of the 
LHC and LC to differentiate new physics associated with KK and $Z'$ 
excitations. In the present study the LC reach was found to be somewhat 
superior to that of the LHC but the LC analysis depended upon the LHC 
determination of the resonance mass as an input. It would be useful to 
perform both of these studies at the level of fast MC to verify the results 
obtained here, including systematic effects as well as the input of 
$A_{FB}$ data at 
the LHC and Bhabha scattering data at the LC. The analysis as presented 
here can also be extended to other scenarios which will be considered 
elsewhere.

As a final point one may wonder about the reverse problem, \ie, if the 
heavy resonance observed at the LHC {is} a $Z'$, how do we rule out the 
possibility of it being a KK state? From performing the analysis discussed 
above at both the LHC and LC we would know that the couplings would be 
consistent with being a $Z'$ and not


\subsection*{Acknowledgements}

The author would like to thank G. Azuelos, J. L. Hewett, and G. Polesello 
for discussions related to this analysis.  


%
\def\MPL #1 #2 #3 {Mod. Phys. Lett. {\bf#1},\ #2 (#3)}
\def\NPB #1 #2 #3 {Nucl. Phys. {\bf#1},\ #2 (#3)}
\def\PLB #1 #2 #3 {Phys. Lett. {\bf#1},\ #2 (#3)}
\def\PR #1 #2 #3 {Phys. Rep. {\bf#1},\ #2 (#3)}
\def\PRD #1 #2 #3 {Phys. Rev. {\bf#1},\ #2 (#3)}
\def\PRL #1 #2 #3 {Phys. Rev. Lett. {\bf#1},\ #2 (#3)}
\def\RMP #1 #2 #3 {Rev. Mod. Phys. {\bf#1},\ #2 (#3)}
\def\NIM #1 #2 #3 {Nuc. Inst. Meth. {\bf#1},\ #2 (#3)}
\def\ZPC #1 #2 #3 {Z. Phys. {\bf#1},\ #2 (#3)}
\def\EJPC #1 #2 #3 {E. Phys. J. {\bf#1},\ #2 (#3)}
\def\IJMP #1 #2 #3 {Int. J. Mod. Phys. {\bf#1},\ #2 (#3)}
\def\JHEP #1 #2 #3 {J. High En. Phys. {\bf#1},\ #2 (#3)}


\begin{thebibliography}{00}

%
\bibitem{anton} 
See, for example,
I. Antoniadis, \PLB B246 377 1990 ;
I. Antoniadis, C. Munoz and M. Quiros, \NPB B397 515 1993 ;
I. Antoniadis and K. Benalki, \PLB B326 69 1994  and \IJMP A15 4237 2000 ;
I. Antoniadis, K. Benalki and M. Quiros, \PLB B331 313 1994 . 
%
\bibitem{bkt}
M.~Carena, E.~Ponton, T.~M.~Tait and C.~E.~Wagner,
arXiv:hep-ph/0212307;
M.~Carena, T.~M.~Tait and C.~E.~Wagner,
Acta Phys.\ Polon.\ B {\bf 33}, 2355 (2002)
[arXiv:hep-ph/0207056];
H.~Davoudiasl, J.~L.~Hewett and T.~G.~Rizzo,
arXiv:hep-ph/0212279.
%
\bibitem{ued}
T.~Appelquist, H.~C.~Cheng and B.~A.~Dobrescu,
Phys.\ Rev.\ D {\bf 64}, 035002 (2001)
[arXiv:hep-ph/0012100]; 
H.~C.~Cheng, K.~T.~Matchev and M.~Schmaltz,
Phys.\ Rev.\ D {\bf 66}, 056006 (2002)
[arXiv:hep-ph/0205314] and 
Phys.\ Rev.\ D {\bf 66}, 036005 (2002)
[arXiv:hep-ph/0204342]; 
T.~G.~Rizzo,
Phys.\ Rev.\ D {\bf 64}, 095010 (2001)
[arXiv:hep-ph/0106336].
%
\bibitem{bunch}
See, for example,  T.G.~Rizzo and J.D.~Wells, \PRD D61 016007 2000 ;
P. Nath and M. Yamaguchi, \PRD D60 116006 1999 ;
M. Masip and A. Pomarol, \PRD D60 096005 1999 ; 
L.~Hall and C.~Kolda, \PLB B459 213 1999 ; 
R. Casalbuoni, S. DeCurtis, D. Dominici and R. Gatto, \PLB B462 48 1999 ;
A.~Strumia, \PLB B466 107 1999 ; 
F. Cornet, M. Relano and J. Rico, \PRD D61 037701 2000 ;
C.D. Carone, \PRD D61 015008 2000 .
%
\bibitem{tgr}
T.G. Rizzo, \PRD D61 055005 2000 ~and \PRD D64 015003 2001 .
%
\bibitem{rs}
L.~Randall and R.~Sundrum,
Phys.\ Rev.\ Lett.\  {\bf 83}, 3370 (1999)
[arXiv:hep-ph/9905221].
%
\bibitem{dhr} 
For an overview of the Randall-Sundrum model phenomenology, see 
H. Davoudiasl, J.L. Hewett and T.G. Rizzo, \PRL 84  2080 2000 ; 
~\PLB B493 135 2000 ;~ and \PRD D63 075004 2001 .
%
\bibitem{snow}
For a review of new gauge boson physics at colliders and details of the 
various models, see 
J.L.\ Hewett and T.G.\ Rizzo, \PR 183 193 1989 ;
M. Cvetic and S. Godfrey, in {\it Electroweak Symmetry Breaking
and Beyond the Standard Model}, ed. T. Barklow \etal, (World Scientific,
Singapore, 1995), hep-ph/9504216; T.G. Rizzo in {\it New Directions for High 
Energy Physics: Snowmass 1996}, ed. D.G. Cassel, L. Trindle Gennari 
and R.H. Siemann, (SLAC, 1997), hep-ph/9612440; 
A. Leike, \PR 317 143 1999  .
%
\bibitem{models}
This is a common feature of the class of models wherein the usual $SU(2)_L$ of 
the SM is the result of a diagonal breaking of a product of two or more 
$SU(2)$'s. For a discussion of a few of these models, see
H. Georgi, E.E. Jenkins, and E.H. Simmons, \PRL 62 2789 1989 ~and 
\NPB B331 541 1990 ; V. Barger and T.G. Rizzo, \PRD D41 946 1990 ;
T.G. Rizzo, \IJMP A7 91 1992 ; R.S. Chivukula, E.H. Simmons and J. Terning, 
\PLB B346 284 1995 ;
A. Bagneid, T.K. Kuo, and N. Nakagawa, \IJMP A2 1327 1987 ~and \IJMP A2 
1351 1987 ; D.J. Muller and S. Nandi, \PLB B383 345 1996 ; 
X.Li and E. Ma, \PRL 47 1788 1981 ~and \PRD D46 1905 1992 ;
E. Malkawi, T.Tait and C.-P. Yuan, \PLB B385 304 1996 .
%
\bibitem{before}
See, for example, the analysis by G. Azuelos and G. Polesello in, 
G.~Azuelos {\it et al.},
``The beyond the standard model working group: Summary report,''
~Proceedings of Workshop on Physics at TeV Colliders, 
Les Houches, France, 21 May - 1 Jun 2001, arXiv:hep-ph/0204031;
T.~G.~Rizzo,
in {\it Proc. of the APS/DPF/DPB Summer Study on the Future of Particle 
Physics (Snowmass 2001) } ed. N.~Graf, eConf {\bf C010630}, P304 (2001)
[arXiv:hep-ph/0109179].
%
\bibitem{upgrade}
F. Gianotti \etal, arXiv:hep-ph/0204087.
%
\bibitem{oldt}
T.~G.~Rizzo,
Phys.\ Rev.\ D {\bf 55}, 5483 (1997)
[arXiv:hep-ph/9612304].
%
\bibitem{oldtt}
T.~G.~Rizzo,
[arXiv:hep-ph/0303056]. 
%
\end{thebibliography}
\end{document}